\documentclass[journal]{IEEEtran}
\usepackage{cite}

\ifCLASSINFOpdf
\usepackage[pdftex]{graphicx}
\else
\fi
\usepackage{amsmath,amssymb}
\usepackage{array}
\usepackage{url}

\begin{document}
%
\title{Excitation of Surface Waves with On-Demand Polarization at Self-Complementary Metasurface}
%
%
%

\author{Sergey~Polevoy,
        and~Oleh~Yermakov
\thanks{This work was supported in part by the National Academy of Science of Ukraine (project "All-dielectric metasurfaces with polarization-degenerate spectrum for high-precision angle-resolved chiral sensing, Agreement 05/01-2023(4) on 02.01.2023), and in part by the Ministry of Education and Science of Ukraine under Grant 0122U001482. }
\thanks{S. Polevoy is with the Radiospectroscopy Department, O.~Ya.~Usikov Institute for Radiophysics and Electronics of the NAS of Ukraine, Kharkiv, Ukraine (e-mail: polevoy@ire.kharkov.ua).}
\thanks{O. Yermakov is with the Department of Computer Physics, V.~N.~Karazin Kharkiv National University, Kharkiv, Ukraine (e-mail: oe.yermakov@gmail.com).}
}

%
%

\markboth{}%
{Polevoy \MakeLowercase{\textit{et al.}}: Excitation of Surface Waves with On-Demand Polarization  at Self-Complementary Metasurface}
%



\maketitle

\begin{abstract}
Surface electromagnetic waves are the main carriers of the highly localized in-plane signal in planar antennas and photonic devices. However, the polarization degree of freedom is generally absent for surface waves as far as it is still a challenge to excite the surface waves with a necessary polarization state. In this letter, for the first time, we propose a method to excite the surface waves of arbitrary polarization using the self-complementary metasurfaces obeying Babinet's duality principle. Namely, we demonstrate the excitation of surface waves with (i) linear horizontal and vertical,  (ii) linear diagonal, (iii) right- and left-handed circular, and (iv) elliptical polarizations covering, in principle, a whole Poincar{\'e} sphere of polarization states. Finally, we analyze the impact of substrate on the efficiency of polarization-demanded surface waves excitation. The results obtained may find the numerous applications in planar photonic and flat optical polarization devices.
\end{abstract}


%
\IEEEpeerreviewmaketitle

\section{Introduction}
%
%
%
%
\IEEEPARstart{P}{lanar} photonic and flat optical devices is a highly promising research topic paving way towards in-plane electromagnetic signal transferring and processing. Control over in-plane information signal requires high field localization referring to surface electromagnetic waves. Specific interest is paid to the surface waves at two-dimensional materials and structures~\cite{maci2011metasurfing,gomez2016flatland,nemilentsau2016anisotropic}. For the last decade, the great success was achieved in aspect of the dispersion engineering and wavefront shaping~\cite{high2015visible,gomez2015hyperbolic,yermakov2015hybrid,samusev2017polarization,li2018infrared,ma2018plane}, high directivity and unidirectional excitation~\cite{pors2014efficient,correas2017plasmon,sinev2017chirality,gangaraj2019unidirectional,nemilentsau2019switchable,meng2020optical,sinev2020steering,coppolaro2021surface} of surface waves at metasurfaces and 2D materials. Despite hybrid TE-TM polarization of surface waves at anisotropic metasurface~\cite{yermakov2016spin,yermakov2018experimental}, the excitation of surface waves with predefined polarization has not been achieved to the best of our knowledge.

To reach the polarization degree of freedom for surface waves, one requires the coincident dispersions of two surface modes with orthogonal (TE and TM) polarizations. The dispersion degeneracy of TE and TM plane waves is the inherent feature of the bulk isotropic medium~\cite{born2013principles}, but it disappears under miniaturization and switching to localized eigenmodes spectrum. This problem may be potentially resolved by the near-field overlapping of electric and magnetic resonances~\cite{yermakov2019broadband,asadulina2021polarization,yermakov2022merging}. 

Another approach is based on the Babinet's duality principle implemented in self-complementary metasurface (SCM)~\cite{nakata2013plane,baena2017broadband}. SCMs obeying Babinet's duality principle are widely used for the manipulation of plane waves~\cite{nakata2013plane,urade2015frequency} such as narrow band-pass filters~\cite{ortiz2013self}, perfect absorbers~\cite{urade2016broadband}, ultrathin polarizers~\cite{baena2015self,baena2017broadband} and beam splitters~\cite{kuznetsov2021self}. Even more intriguing are the properties of surface waves supported by the SCMs including all-frequency hyperbolicity, extreme canalization and degenerate TE-TM dispersion in the special directions~\cite{yermakov2021surface}. The latter brings the near-field polarization degree of freedom, which is used within this work.


In this work, we study numerically the near-field properties of self-complementary metasurfaces, namely, we investigate the excitation of surface waves. Using the above-mentioned polarization degeneracy of TE and TM surface modes along the specific directions, we aim exciting the surface waves of the selected polarization on demand. We show that it may be realized by adjusting the mutual amplitudes and phases of two sets of the excitation ports. First, we demonstrate the independent excitation of TE- and TM-polarized surface waves (Section~\ref{sec-S1}). Then, using the superposition of TE and TM eigenmodes, we demonstrate the excitation of surface waves with linear diagonal and circular polarizations (Section~\ref{sec-S23}), as well as with on-demand elliptical polarization (Section~\ref{sec-S123}). Finally, we analyze the impact of the substrate on the degree of surface waves polarization (Section~\ref{sec:subs}). The results obtained discover the near-field polarization degree of freedom for surface waves.

\section{Design and Methods}

\subsection{Main Operational Principle}

Following the effective local medium approximation one describes the SCM via the two-dimensional effective surface impedance~\cite{baena2017broadband} or admittance~\cite{yermakov2021surface} tensor. It has been recently shown that the dispersions of TE and TM surface waves at mutually complementary metasurfaces of capacitive- and inductive-type are equal to each other in a free space~\cite{gonzalez2014surface,dia2017guiding}. In particular, this principle is used for the effective guiding of surface wave via a topological state~\cite{dia2017guiding,bisharat2019electromagnetic}. For the SCM, the dispersion of TE and TM surface waves are completely degenerate only along the main axes of the surface admittance tensor and defined as follows~\cite{yermakov2021surface}:
\begin{equation}
\begin{split}
    & k_x = \Theta[\text{Im}(Y_x)] \, k_0 \sqrt{1 - Y_y/Y_x},\\
    & k_y = \Theta[\text{Im}(Y_y)] \, k_0 \sqrt{1 - Y_x/Y_y},
\end{split}
\label{disp_k}
\end{equation}
where $k_{x,y}$ are the wave vectors of surface waves along $x$- and $y$-direction, respectively, $Y_{x,y}$ are the corresponding components of the effective surface admittance tensor, $\Theta[\,]$ is the Heaviside unit step function, $k_0 = 2 \pi f /c = 2\pi/\lambda$ is the wave vector in vacuum, $f$ is the operational frequency, $\lambda$ is the operational wavelength, $c$ is the speed of light.

Let us assume the propagation of two degenerate modes along $x$-direction. Then, the broadband degeneracy of orthogonally polarized TE ($E_y, H_x, H_z$) and TM ($E_x, E_z, H_y$) eigenmodes means that one can implement their superposition at the same frequency:
\begin{equation}
\begin{split}
    &\mathbf{E} = \left( A_p E_x, A_s E_y, A_p E_z \right) e^{i k_x x - i \omega t},\\
    &\mathbf{H} = \left( A_s H_x, A_p H_y, A_s H_z \right) e^{i k_x x - i \omega t},
\end{split}
\end{equation}
where $A_s$ and $A_p$ are the complex amplitudes of TE and TM eigenmodes, respectively. By adjusting appropriately the complex amplitude values ratio $A_s/A_p$, i.e. their absolute values ratio and phase difference, one can achieve any polarization state on demand covering completely a Poincar{\'e} sphere. Practically, we cannot set explicitly $A_s$ and $A_p$ complex amplitudes of eigenmodes, but it is possible to control the complex amplitudes of Port 1 ($P_s$) and Port 2 ($P_p$).



\subsection{Design of Self-Complementary Metasurface}

\begin{figure}[t] 
    \centering
    \includegraphics[width=0.99\linewidth]{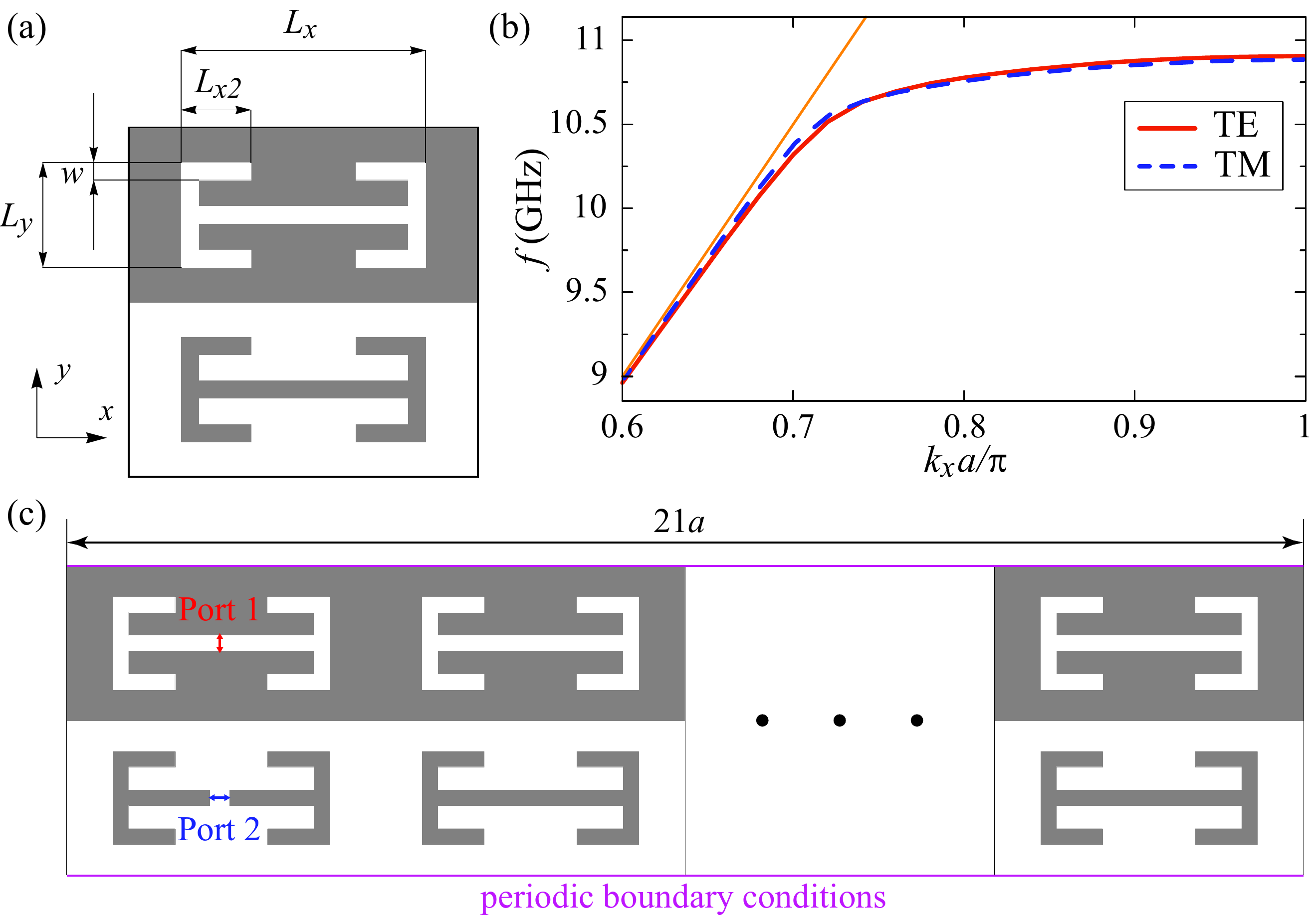}
    \caption{(a) Unit cell of the self-complementary metasurface under study. The metal areas are shown in gray color. (b) Dispersion diagram $f(k_x)$ along $x$-direction for first two modes (TE and TM) of self-complementary metasurface shown in (a). $k_x a /\pi = 1$ corresponds to the boundary of the first Brillouin zone. Orange line corresponds to the light line in vacuum. (c) Simulation model representing a chain of 21 unit cells along $x$-direction with applied periodic boundary conditions along $y$-axis. The surface waves are excited by Port~1 and Port~2 located within the first unit cell (red and blue arrows).}
    \label{fig-geom}
\end{figure}

The unit cell of the self-complementary metasurface under study is shown in Fig.~\ref{fig-geom}a. It consists of two complementary sections: I-shaped metal element and inverse hole in the metal screen. Period and thickness of metasurface are $a = 10$~mm and $h_m=0.0035a$, respectively. The parameters of each I-shaped element are shown in Fig.~\ref{fig-geom}a and defined as follows $L_x=0.7a$, $L_{x2}=0.2a$, $L_y=0.3a$, $w=0.05a$. We assume a metasurface is located at the substrate with thickness $h=0.1a$, permittivity $\varepsilon=1$ and loss tangent $\text{tan}\delta = 0.02$, which emulates the possible roughness and defects losses. In Section~\ref{sec:subs} we analyze the case with a real FR-4 substrate ($\varepsilon = 3.9$, $\text{tan}\delta = 0.02$)~\cite{edwards2016foundations} of the same thickness.

As it was previously discussed, according to Eq.~\eqref{disp_k} the dispersions of TE and TM surface waves at SCM are degenerate along $x$-direction (Fig.~\ref{fig-geom}b). The numerical simulation of the dispersion diagram has been done in Eigenmode Solver of CST Microwave Studio. The small discrepancy of a complete degeneracy is associated with the finite thickness of a metal layer and the loss tangent of the substrate.



Since the first two modes are orthogonal with respect to polarization and rather strongly degenerate, we use this metasurface as a platform for the excitation of surface waves with specified polarization. The finite-size and full-scale simulation model consists of 21 unit cells along $x$-axis. The periodic boundary conditions are applied along the $y$-axis while we consider the wave propagation along $x$-axis (Fig.~\ref{fig-geom}c). The total simulated area in $xz$-plane is equal to 230$\times$137.5~mm$^2$. At left side of the structure, two orthogonally-directed electric dipoles are placed in the first unit cell. Port~1 is placed in the existing hole, while we added a gap in metal for Port~2 of thickness $w$ (Fig.~\ref{fig-geom}c). One should note that we consider a two sets of ports indeed as far as the periodic boundary conditions are applied. The amplitude and phase of each port may change that makes it possible to emit a surface wave with the desired polarization. All the results presented are verified by the full-scale numerical simulation performed in the frequency domain using the finite-element method applying the Floquet-type periodic boundary conditions. To analyze the proper state of polarization, we calculate the $E_y$ and $E_z$ components of electric field in each point of $xz$-plane at $y=0$ with a step of 1~mm.

\section{Results and Discussion}

In this Section, we analyze the spatial distribution of the electric field via the Stokes parameters ($S_0, S_1, S_2, S_3$) formalism~\cite{born2013principles}. Assuming wave propagation along $x$-axis, we define the Stokes parameters as follows $S_0 = |E_y|^2 + |E_z|^2$, $S_1 = |E_y|^2 - |E_z|^2$, $S_2 = 2\, \text{Re}(E_y E_z^*)$ and $S_3 = -2\, \text{Im}(E_y E_z^*)$. Namely, we show the degrees of linear horizontal and vertical ($S_1$), linear rotated at 45$^\circ$ and 135$^\circ$ with respect to $y$-axis ($S_2$), right- and left-handed circular ($S_3$) polarizations normalized per total intensity ($S_0$). The typical values of Stokes parameters are shown in Table~\ref{table-stokes}. Here, we define the right- and left-handed circular or elliptic polarizations as a clockwise and counter-clockwise rotation with respect to the surface wave propagation ($x$-axis), respectively. 
\begin{table}[htb]
\centering
\caption{The typical wave polarizations and corresponding values of Stokes parameters.}
\begin{tabular}{ m{2.7cm}    m{2.2 cm}  }
\hline
Degree of polarization & Stokes parameters \\ \hline 
linear horizontal & $S_1/S_0 = 1$ \\ 
linear vertical & $S_1/S_0 = -1$ \\ 
linear at 45$^\circ$ & $S_2/S_0 = 1$ \\ 
linear at 135$^\circ$ & $S_2/S_0 = -1$ \\ 
left-handed circular & $S_3/S_0 = 1$ \\ 
right-handed circular & $S_3/S_0 = -1$ \\  \hline 
\end{tabular}
\label{table-stokes}
\end{table}

\subsection{Excitation of Surface Waves by Single Port: Linear TE and TM Polarizations \label{sec-S1}}

\begin{figure}[h] 
    \centering
    \includegraphics[width=0.63\linewidth]{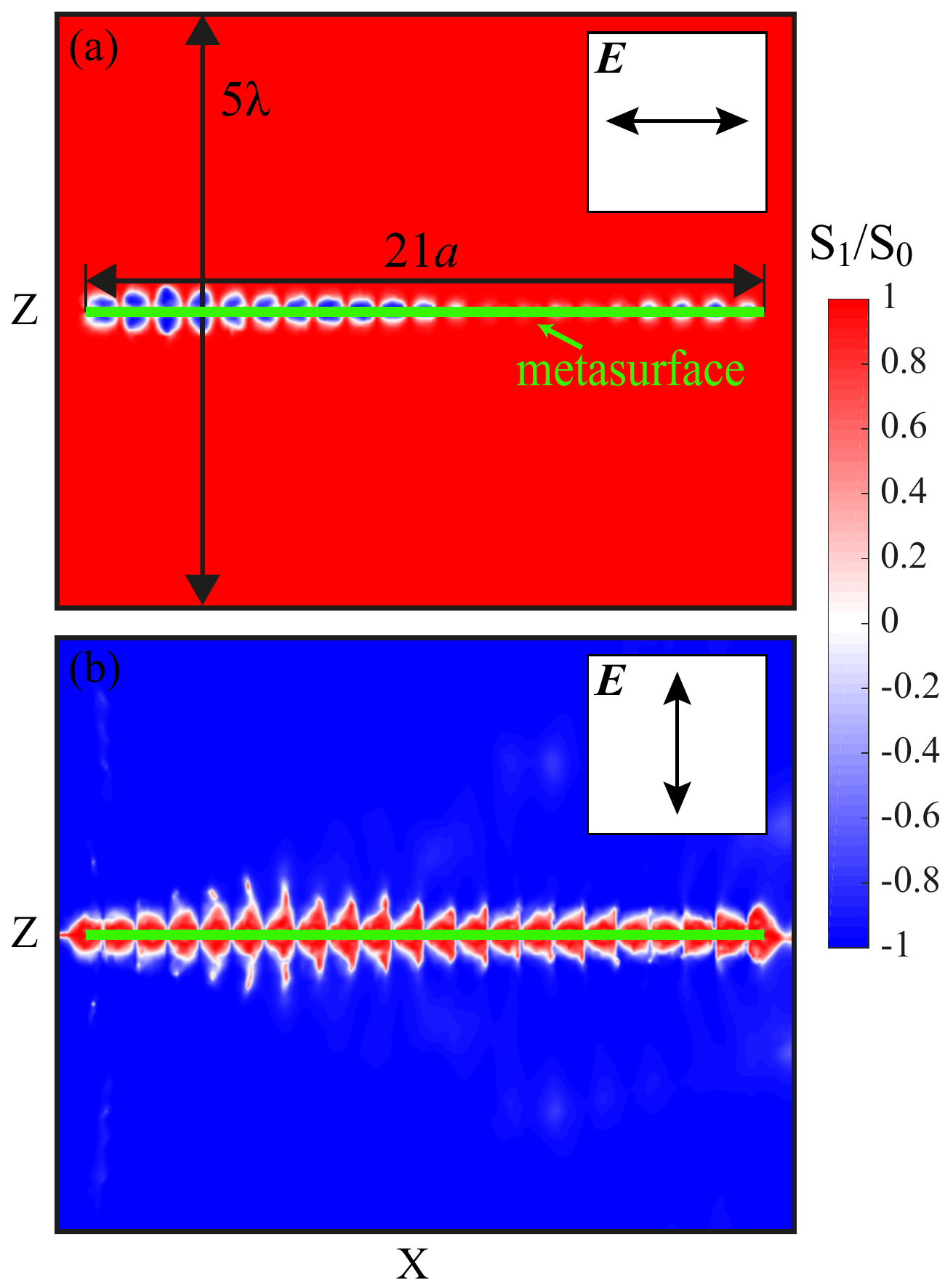}
    \caption{Spatial distribution of $S_1$ Stokes parameter at $f = 10.9$~GHz in $xz$-plane at $y=0$ under the excitation of (a) Port~1 ($P_s = 1$) and (b) Port~2 ($P_p = 1$). Green bar corresponds to the arrangement of self-complementary metasurface consisting of 21 unit cells along $x$-axis. We consider the area of 230 mm$\times$137.5~mm (8.4$\lambda \times$5$\lambda$). The color bar shows the degree of $S_1/S_0$ in units from $-1$ (vertical linear polarization) to $1$ (horizontal linear polarization). }
    \label{fig-S1}
\end{figure}

First, we consider the excitation only by a single port. The excitation by the Port~1 ($P_s = 1, P_p =0$) or Port~2 ($P_s = 0, P_p =1$) results in the excitation of TE or TM surface wave, respectively (Fig.~\ref{fig-S1}). The polarization of surface wave in any point except the vicinity of metasurface is almost purely linear horizontal (TE) or vertical (TM). The typical values of $|S_1|$ exceed 0.97 and 0.94 at the distance $|z| > 7$~mm (about $\lambda/4$) and $|z| > 14$~mm (about $\lambda/2$) under excitation by Port~1 and Port~2, respectively. So, we demonstrate the independent and controlled excitation of TE and TM surface waves in both near- and far-field. 

\subsection{Excitation of Surface Waves by Two Ports: Linear under $\pm 45^\circ$ and Left/Right-Handed Circular Polarizations \label{sec-S23}}

Now, we can implement the excitation of surface wave with the desired polarization as the superposition of TE and TM modes. For this reason, we optimize numerically the parameters of the ports, namely their absolute values ratio and phase difference. 

For further analysis, we introduce the averaged value of Stokes parameters over the chosen area defined as follows:
\begin{equation}
    \overline{S}_i = \dfrac{\int S_i \, dx dz}{\int {dx dz}}, \; i = 1,2,3.
\end{equation}

\begin{figure}[t] 
    \centering
    \includegraphics[width=0.99\linewidth]{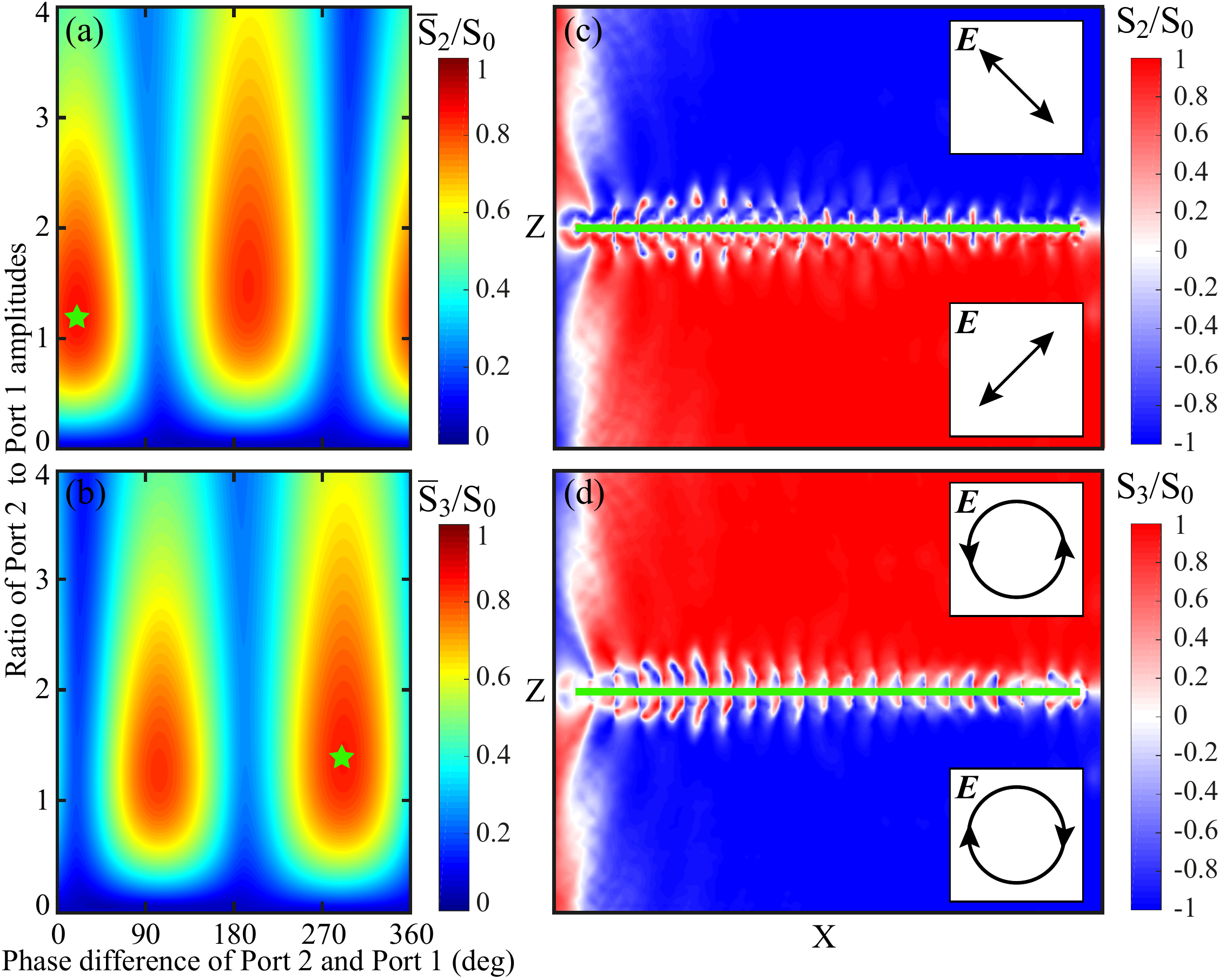}
    \caption{(a,b) The dependence of average (a) $|\overline{S}_2|/S_0$ and (b) $|\overline{S}_3|/S_0$ Stokes parameters on the ports complex amplitudes. Here, we take $P_s = 1$ and change modulus $|P_p|$ and phase [$\text{arg}(P_p)$] of the second port. The color bar shows the values of $|\overline{S}_2|/S_0$ and $|\overline{S}_3|/S_0$ from 0 (blue) to 1 (red). (c,d) Spatial distribution of (c) $S_2$ and (d) $S_3$ Stokes parameters at $f = 10.9$~GHz in $xz$-plane at $y=0$ under the excitation ports parameters marked by the green stars in (a) $P_p/P_s = 1.2$, $\text{arg}(P_p) - \text{arg}(P_s) = 20^\circ$ and (b) $P_p/P_s = 1.4$, $\text{arg}(P_p) - \text{arg}(P_s) = -70^\circ$, respectively. The color bars show the degree of (c) $S_2/S_0$ in units from $-1$ (linear polarization rotated at 135$^\circ$) to $1$ (linear polarization rotated at 45$^\circ$), and (d) $S_3/S_0$ in units from $-1$ (right-handed circular polarization) to $1$ (left-handed circular polarization). }
    \label{fig-S2-S3}
\end{figure}

Then, we fix the complex amplitude of Port~1 as $P_s = 1$ and change absolute value $|P_p|$ and phase [$\text{arg}(P_p)$] of Port~2 in order to find the maxima of the necessary Stokes parameters (marked by the green stars in Figs.~\ref{fig-S2-S3}a-\ref{fig-S2-S3}b).

Figures~\ref{fig-S2-S3}c and \ref{fig-S2-S3}d demonstrate the distribution of $S_2$ and $S_3$ Stokes parameters, respectively. One can notice the near-perfect excitation of desired polarizations. The averaged value of $\overline{S}_2$ and $\overline{S}_3$ Stokes parameters are 0.86 and 0.85, respectively. Considering the region after first 5 periods of metasurface from the excitation port side at the height of $|z| > 21$~mm (about $3\lambda/4$), we find the $|S_2|$ and $|S_3|$ Stokes parameters exceed 0.94.

\subsection{Excitation of Surface Waves by Two Ports: Arbitrary Elliptic Polarization \label{sec-S123}}

\begin{figure}[t] 
    \centering
    \includegraphics[width=0.95\linewidth]{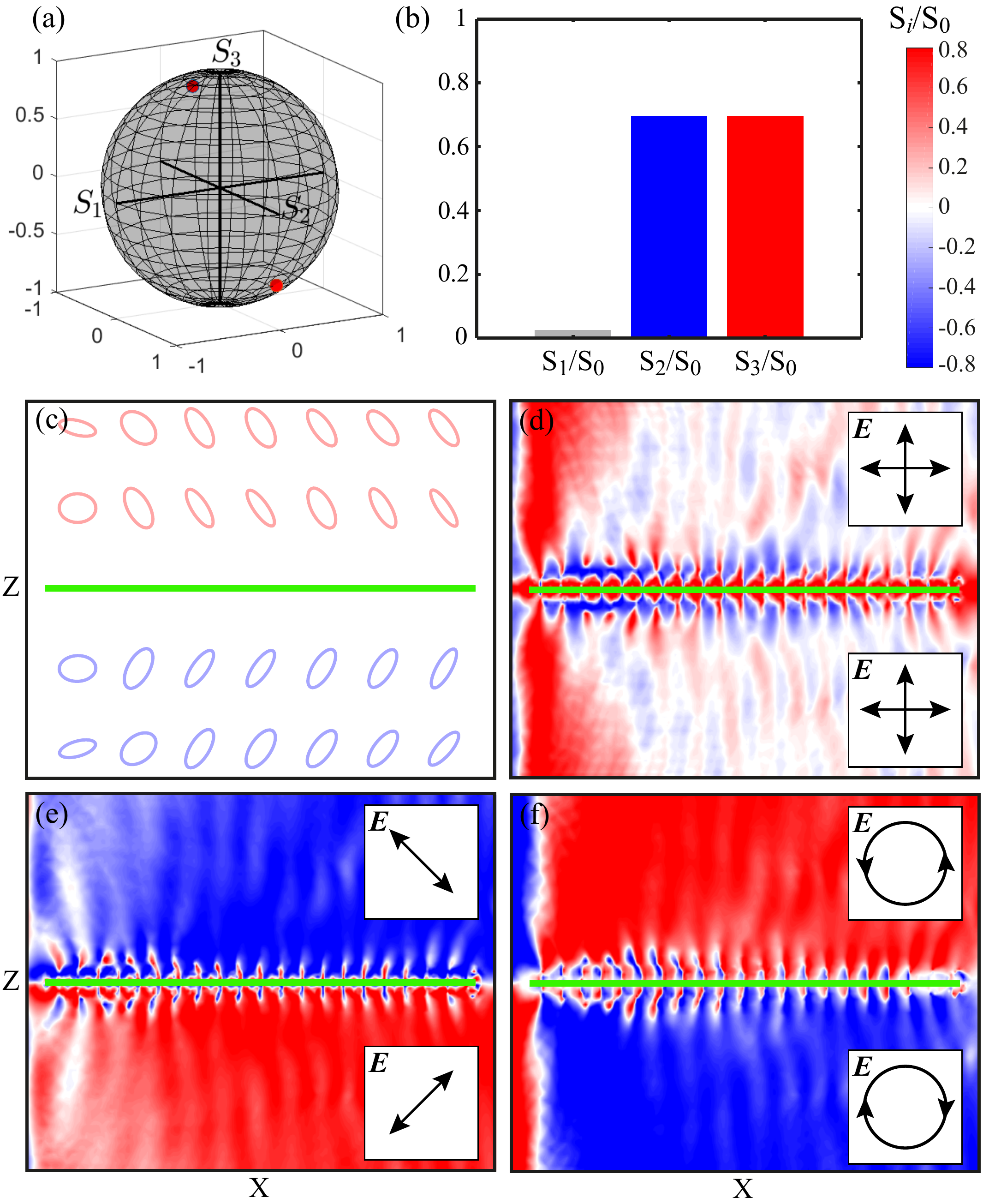}
    \caption{(a) Poincar{\'e} sphere with labeled states of polarization marked by red points ($S_1/S_0 = 0.02, S_2/S_0 = \pm 0.697, S_3/S_0 = \mp 0.697$). (b) The bar diagrams of the Stokes parameters labeled in (a). (c) The elliptical polarization states in several points at the heights $z = \pm \lambda$ and $z = \pm 2 \lambda$. Red and blue colors of ellipses correspond to the counter- and clockwise rotation directions, respectively. (d-f) Spatial distribution of (d) $S_1/S_0$, (e) $S_2/S_0$ and (f) $S_3/S_0$ Stokes parameters at $f = 10.9$~GHz in $xz$-plane at $y=0$ under the excitation ports parameters $P_p/P_s = 1.2$, $\text{arg}(P_p) - \text{arg}(P_s) = -25^\circ$. The uniform color bar shows the degree of each Stokes parameter $S_i/S_0$, where $i=1,2,3$.}
    \label{fig-elliptic}
\end{figure}

Finally, we demonstrate the excitation of surface wave with arbitrary elliptic polarization. We considered the smaller region above the metasurface, namely above 18~mm and 45~mm to the right of the port. We choose the elliptical polarization state with $\overline{S}_1 = 0.02$, $\overline{S}_2 = \mp 0.697, \overline{S}_3 = \pm 0.697$ (Figs.~\ref{fig-elliptic}a-\ref{fig-elliptic}b) in this region, where the upper and lower signs correspond to the areas above and below a metasurface, respectively. The corresponding parameters of Port~2 are $P_p = 1.2$, $\text{arg}(P_p) = -25^\circ$, while $P_s = 1$. The related spatial distributions of $S_1$, $S_2$ and $S_3$ Stokes parameters are shown in Figs.~\ref{fig-elliptic}d, \ref{fig-elliptic}e and \ref{fig-elliptic}f, respectively. The distribution of $S_1$ parameter is at the level of noise with values several times smaller than in the case of $S_2$ and $S_3$. The respective polarization ellipses are drawn in several points at the heights $z=\pm \lambda$ and $z=\pm 2\lambda$ in Fig.~\ref{fig-elliptic}c.

Indeed, almost any polarization state of surface wave may be achieved using the proper parameters of the ports, covering a Poincar{\'e} sphere completely.

\subsection{Effect of Substrate \label{sec:subs}}

\begin{figure}[h] 
    \centering
    \includegraphics[width=0.99\linewidth]{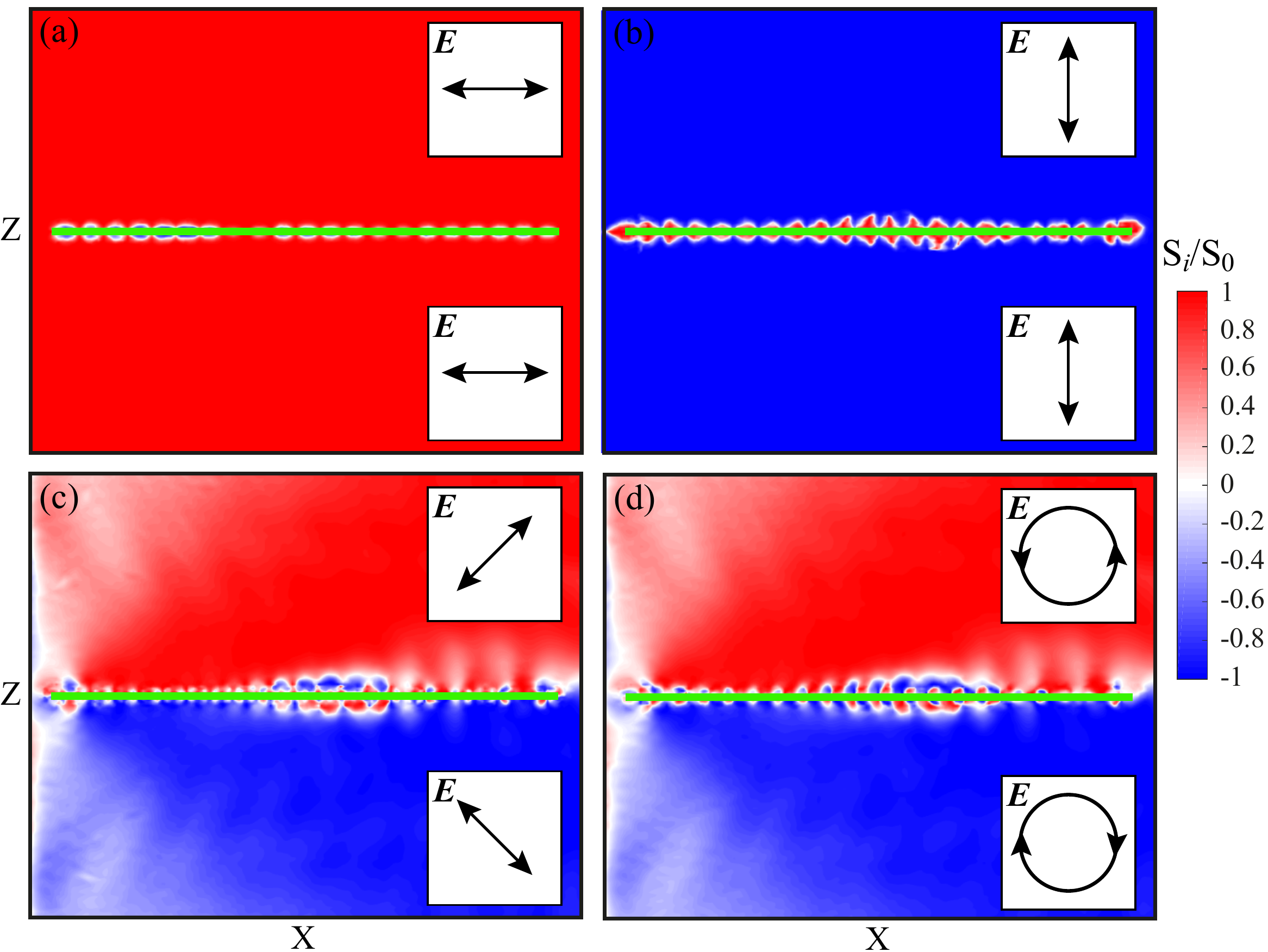}
    \caption{Spatial distribution of Stokes parameters (a,b) under single port excitation ($S_1$) and (c,d) under two ports excitation ($S_2$, $S_3$) in the case of self-complementary metasurface on FR-4 substrate at $f=7.2$~GHz in $xz$-plane. We consider the area of 230 mm$\times$210~mm (5.5$\lambda \times$5$\lambda$). The ports parameters are (c) $P_p/P_s = 0.8$, $\text{arg}(P_p) - \text{arg}(P_s) = 144^\circ$ and (d) $P_p/P_s = 0.8$, $\text{arg}(P_p) - \text{arg}(P_s) = -123^\circ$. The uniform color bar shows the degree of each Stokes parameter $S_i/S_0$, where $i=1,2,3$. }
    \label{fig-subs}
\end{figure}

From practical point of view and real implementation, it is important to analyze the impact of the substrate on the polarization-specified excitation of surface waves. For this reason, we use the popular microwave substrate FR-4 ($\varepsilon = 3.9$, $\text{tan}\delta = 0.02$)~\cite{edwards2016foundations}. One should note that, strictly speaking, the Babinet's principle is slightly violated in this case~\cite{gonzalez2014surface}. Nevertheless, it does not lead to the significant discrepancy in terms of polarization-designated excitation of surface waves as one can see in Fig.~\ref{fig-subs}.

Figures~\ref{fig-subs}c and \ref{fig-subs}d show the distribution of $S_2$ and $S_3$ Stokes parameters, respectively. Considering the region after first 5 periods of metasurface from the excitation port side at the height of $|z| > 32$~mm (about $3\lambda/4$), we find the $|S_2|$ and $|S_3|$ Stokes parameters exceed 0.89.

\newpage
\section{Conclusion}
We have examined the excitation of surface waves with predefined polarization using the unique properties of self-complementary metasurfaces. In particular, we have demonstrated the excitation of surface waves with TE and TM, linear under $\pm45^\circ$, right- and left-handed circular and elliptical polarization, proving the ability to achieve any polarization state on demand. We have shown the impact of substrate bringing negligible contribution, not affecting the principal results of this work and, as a consequence, opening way for the experimental verification. The results obtained discover a new degree of freedom for the near-field polarization control enriching the fundamental spin-orbit phenomena and practical applications in planar photonics and flat optics.


%



\section*{Acknowledgment}


The authors would like to thank the Armed Forces of Ukraine for their service providing the opportunities to perform this research in Ukraine.

\ifCLASSOPTIONcaptionsoff
  \newpage
\fi



%

\newpage
\bibliographystyle{IEEEtran}
\bibliography{references.bib}

%





\end{document}